\documentclass[review]{elsarticle}
\usepackage[utf8]{inputenc}
\usepackage{lineno,hyperref}
\usepackage{tabularx}
\usepackage{graphicx}
\usepackage{adjustbox}
\usepackage{booktabs}
\usepackage{caption}
\usepackage{pdfpages}
\usepackage{float}
\usepackage{colortbl}
\usepackage{rotating}
\usepackage{lscape}
\usepackage{amsmath}
\usepackage{amssymb}
\usepackage{wrapfig}
\usepackage{epstopdf}

\journal{Computers in Biology and Medicine}

\begin{document}

\begin{frontmatter}

\title{Cell segmentation from telecentric bright-field transmitted light microscopy images using a Residual Attention U-Net: a case study on HeLa line}

\author[1]{Ali Ghaznavi}
\author[1]{Renata Rycht\'{a}rikov\'{a}}
\ead{rrychtarikova@frov.jcu.cz}
\author[2]{Mohammadmehdi Saberioon}
\author[1]{Dalibor \v{S}tys}




\address[1]{Faculty of Fisheries and Protection of Waters, South Bohemian Research Center of Aquaculture and Biodiversity of Hydrocenoses, Institute of Complex Systems, University of South Bohemia in \v{C}esk\'{e} Bud\v{e}jovice, Z\'{a}mek 136, 373 33 Nov\'{e} Hrady, Czech Republic}

\address[2]{Helmholtz Centre Potsdam GFZ German Research Centre for Geosciences, Section 1.4 Remote Sensing and Geoinformatics, Telegrafenberg, Potsdam 14473, Germany}



\cortext[mycorrespondingauthor]{Corresponding author: Renata Rycht\'{a}rikov\'{a}}


\begin{abstract}
Living cell segmentation from bright-field light microscopy images is challenging due to the image complexity and temporal changes in the living cells. Recently developed deep learning (DL)-based methods became popular in medical and microscopy image segmentation tasks due to their success and promising outcomes. The main objective of this paper is to develop a deep learning, U-Net-based method to segment the living cells of the HeLa line in bright-field transmitted light microscopy. To find the most suitable architecture for our datasets, a residual attention U-Net was proposed and compared with an attention and a simple U-Net architecture. 

The attention mechanism highlights the remarkable features and suppresses activations in the irrelevant image regions. The residual mechanism overcomes with vanishing gradient problem. The Mean-IoU score for our datasets reaches 0.9505, 0.9524, and 0.9530 for the simple, attention, and residual attention U-Net, respectively. The most accurate semantic segmentation results was achieved in the Mean-IoU and Dice metrics by applying the residual and attention mechanisms together. The watershed method applied to this best -- Residual Attention -- semantic segmentation result gave the segmentation with the specific information for each cell.

\end{abstract}

\begin{keyword}
\texttt{Deep learning \sep Neural network \sep Cell detection \sep Microscopy image segmentation \sep Tissue segmentation \sep Semantic segmentation \sep Watershed segmentation} 
\end{keyword}

\end{frontmatter}


\section{Introduction}
Image object detection and segmentation can be defined as a procedure to localize a region of interest (ROI) in an image and separate an image foreground from its background using image processing and/or machine learning approaches. Cell detection and segmentation are the primary and critical steps in microscopy image analysis. These processes play an important role in estimating the number of the cells, initializing cell segmentation, tracking, and extracting features necessary for further analysis. In the text below, the segmentation methods were categorized as 1) traditional, feature- and machine learning (ML)-based methods and 2) deep learning (DL)-based methods.

\subsection{Traditional cell segmentation methods}
Traditional segmentation methods have achieved impressive results in cell boundary detection and segmentation, with an efficient processing time~\cite{Rojas-Moraleda2017,Tang2015}. These methods include low-level pixel processing approaches. The region-based methods are more robust than the threshold-based segmentation methods~\cite{Tang2015}. However, in low-contrast images, cells placed close together or flat cell regions can be segmented as blobs. Rojas-Moraleda et al.~\cite{Rojas-Moraleda2017} proposed a region-based method on the principles of persistent homology with an overall accuracy of 94.5\%. The iterative morphological and Ultimate Erosion \cite{Wang2016, Fan2013} suffer from poor segment performance when facing small and low-contrast objects.
Guan et al.~\cite{Guan2011} detected rough circular cell boundaries using the Hough transform and the exact cell boundaries using fuzzy curve tracing. Compared with the watershed-based method~\cite{zhou2009}, this method was more robust to the noise and the uneven brightness in the cells. Winter et al.~\cite{Winter2019} combined the image Euclidean distance transformation with the Gaussian mixture model to detect elliptical cells. This method requires solid objects for computing the distance transform. The target objects' large holes or extreme internal irregularities make the distance transform unreliable and reduce the method performance.
Buggenthin et al. \cite{Buggenthin2013} identified nearly all cell bodies and segmented multiple cells instantly in bright-field time-lapse microscopy images by a fast, automatic method combining the Maximally Stable Extremal Regions (MSER) with the watershed method. The main challenges for this method remain the oversegmentation and poor performance for out-of-focus images.

The machine learning methods have expanded due to the microscopy images' complexity and the previous methods' low performance to detect and segment cells. The ML methods can be classified into two groups: supervised vs unsupervised. The supervised methods produce a mathematical function or model from the training data to map a new data sample~\cite{Stuart2010}.
Mualla et al.~\cite{MuallaF2013} utilized the Scale Invariant Feature Transform (SIFT) as a feature extractor and the Balanced Random Forest as a classifier to calculate the descriptive cell keypoints.
The SIFT descriptors were invariant to illumination conditions, cell size, and orientation. 
Tikkanen et al. \cite{Tikkanen2015} developed a method based on the Histogram of Oriented Gradients (HOG) and the Support Vector Machine (SVM) to extract feature descriptors and classify them as a cell or a non-cell in bright-field microscopy data. The proposed method is susceptible to the number of iterations in the training process as a crucial step to eliminating false positive detections.

The unsupervised ML algorithms require no pre-assigned labels or scores for the training data~\cite{Hinton1999}. The best known unsupervised methods are clustering methods.
Mualla et al.~\cite{MuallalF2014-c} segmented unstained cells in bright-field micrographs using a combination of a SIFT to extract key points, a self-labelling, and two clustering methods. This method is fast and accurate but sensitive to the feature selection step to avoid overfitting.

\subsection{Deep Learning cell segmentation methods}
In the last decade, Deep Learning has emerged as a new area of machine learning. The DL methods contain a class of ML techniques that exploit many layers of non-linear information processing for supervised or unsupervised feature extraction and transformation for pattern analysis and classification. The Deep Convolutional Networks exhibited impressive performance in many visual recognition tasks \cite{Girshick2014}. Song et al.~\cite{Song2016} used a multiscale convolutional network (MSCN) to extract scale-invariant features and graph-partitioning method for accurate segmentation of cervical cytoplasm and nuclei. This method significantly improved the Dice metric and standard deviation compared with similar methods.
Shibuya et al. \cite{Shibuya2021} proposed the Feedback U-Net using the convolutional Long Short-Term Memory (LSTM) network for cell image segmentation, working on four classes of \textit{Drosophila} cell image dataset. However, the proposed method suffered from a low accuracy rate depending on the segmented class. Thi et al. \cite{Thi2022} proposed a convolutional blur attention (CBA) network. The network consists of down- and upsampling procedures for nuclei segmentation in standard challenge datasets~\cite{MoNuSeg,DSB}. The authors achieved a good value of the aggregated Jaccard index. The reduced number of trainable parameters led to a reasonable decrease in the computational cost.
Xing et al.~\cite{Xing2016} also proposed an automated nucleus segmentation method based on a deep convolutional neural network (DCNN) to generate a probability map. However, the proposed mitosis counting remains laborious and subjective to the observer.


One of the most popular models for semantic segmentation is Fully Convolutional Network (FCN) architectures. The FCN combines deep semantic information with a shallow appearance to achieve satisfactory segmentation results.
The convolutional networks can take the arbitrary size of input images to train end-to-end, pixel-to-pixel, and produce an output of the corresponding size with efficient inference and learning to achieve semantic segmentation in complex images, including microscopy and medical images \cite{Long2015,Ben-Cohen2016}.
Ronnenberger et al.~\cite{Ronneberger2015} proposed a training strategy that relies on the strong use of data augmentation by applying U-Net Neural Network, contracting the path to capture context, and expanding the path symmetrically to achieve a precise localization. This method was optimized with a low amount of training labelled samples and efficiently performed electron microscopy image segmentation. Long et al. \cite{Long2020} proposed an enhanced U-Net-based architecture called light-weighted U-Net (U-Net+) with a modified encoded branch for potential low-resources computing of nuclei segmentation in bright-field, dark-field, and fluorescence microscopy images. However, the proposed method did not achieve higher accuracy in the Mean-IoU metric. Bagyaraj et al. \cite{Bagyaraj2021} proposed two automatic deep learning networks called U‐Net‐based deep convolution network and U‐Net with a dense convolutional network (DenseNet) for segmentation and detection of brain tumour cells. The authors achieved remarkable results by applying the DenseNet architecture.

As described above, traditional ML methods are not much efficient to segment cells in a microscopy image with a complex background, particularly bright-field microscopy tiny cells \cite{Buggenthin2013, Tikkanen2015, MuallalF2014-c}. These methods cannot build sufficient models for big datasets. On the other hand, some Convolution Neural Networks (CNNs) require a vast number of manually labelled training datasets and higher computational costs compared with the ML methods~\cite{Long2015,Liu2015}. 

Deep learning-based methods have delivered better outcomes in segmentation tasks than other methods. Therefore, the main objective of this research is to propose a highly accurate and reasonably computationally cost deep learning-based method to segment human HeLa cells in unique telecentric bright-field transmitted light microscopy images. The U-Net was chosen since it is one of the most promising methods used in semantic segmentation~\cite{Ronneberger2015}. Different U-Net architectures such as Attention and Residual Attention U-Net were examined to find the most suitable architecture for our datasets.

Human Negroid cervical epithelioid carcinoma line HeLa~\cite{Lyapun2019} was chosen as a testing cell line for described microscopy image segmentation. The reason for choosing is that HeLa is the oldest, immortal, and most used model cell line ever. HeLa is cultivated in almost all tissue and cell laboratories worldwide and utilized in many fields of medical research, such as research on carcinoma or testing the material biocompatibility.

The processed microscopy data are specific to high-pixel resolution in rgb mode and requires preprocessing to suppress optical vignetting and camera noise. The data shows unlabelled living cells in their physiological state. The cells are shown in-focused and out-of-focus. Thus, the obtained segmentation method is applicable in a 3D visualization of the cell.

\section{Materials and methods}
\subsection{Cell preparation and microscope specification} \label{microscopy}
Human HeLa cell line (European Collection of Cell Cultures, Cat. No. 93021013) was cultivated to low optical density overnight at 37$^{\circ}$C, 5\% CO$_2$, and 90\% relative humidity. The nutrient solution consisted of Dulbecco's modified Eagle medium (87.7\%) with high glucose ($>$1 g L$^{-1}$), fetal bovine serum (10\%), antibiotics and antimycotics (1\%), L-glutamine (1\%), and gentamicin (0.3\%; all purchased from Biowest, Nuaille, France). The HeLa cells were maintained in a Petri dish with a cover glass bottom and lid at room temperature of 37$^{\circ}$C.

Time-lapse image series of living human HeLa cells on the glass Petri dish were captured using a high-resolved bright-field light microscope for observation of microscopic objects and cells. This microscope was designed by the Institute of Complex System (ICS, Nov\'{e} Hrady, Czech Republic) and built by Optax (Prague, Czech Republic) and ImageCode (Brloh, Czech Republic) in 2021. The microscope has a simple construction of the optical path. The light from two light-emitting diods CL-41 (Optika Microscopes, Ponteranica, Italy) passes through a sample to reach a telecentric measurement objective TO4.5/43.4-48-F-WN (Vision \& Control GmbH, Shul, Germany) and an Arducam AR1820HS 1/2.3-inch 10-bit RGB camera with a chip of 4912$\times$3684 pixel resolution. The images were captured as a primary (raw) signal with theoretical pixel size (size of the object projected onto the camera pixel) of 113 nm. The software (developed by the ICS) controls the capture of the primary signal with the camera exposure of 2.75 ms. All these experiments were performed in time-lapse to observe cells' behaviour over time.

\subsection{Data acquisition}\label{dataprep}
Different time-lapse experiments on the HeLa cells were completed under the bright-field microscope (Sect.~\ref{microscopy}). The algorithm proposed in~\cite{Platonova2021} was fully automated and implemented in the microscope control software to calibrate the microscope optical path and correct all image series to avoid image background inhomogeneities and noise. 

After the image calibration, we converted the raw image representations to 8-bit colour (rgb) images of resolution (number of pixels) quarter of the original raw images. We employed quadruplets of Bayer mask pixels \cite{Stys2016}: Red and blue camera filter pixels were adopted into the relevant image channel and each pair of green camera filter pixels' intensities were averaged to create the green image channel. Then, images were rescaled to 8-bits after creating the image series intensity histogram and omitting unoccupied intensity levels. This bit reduction ensured the maximal information preservation and mutual comparability of the images through the time-lapse series.

The means denoising method \cite{Buades2005} minimized the background noise in the constructed RGB images at preserving the texture details. Afterwards, the image series were cropped to the $1024\times1024$ pixel size. The steps described above gave us 500 images from different time-lapse experiments. The image dataset is accessible at the Dryad~\cite{Dryad}. 

\begin{figure}
\graphicspath{ {./images/} }
    \centering
   \includegraphics[width=.9\textwidth]{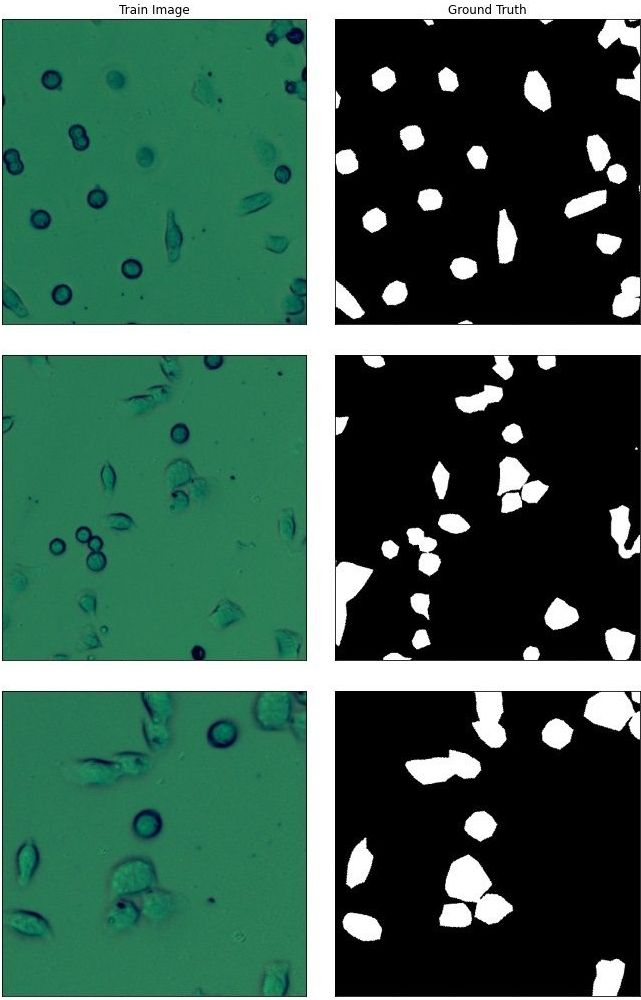}
    \caption{Examples of the train sets and their ground truths. The image size is $512\times512$.} \label{fig1}
\end{figure}

The cells in the images were labelled manually by MATLAB (MathWorks Inc., Natick, Massachusetts, USA) as Ground-Truth (GT) single class masks with the dimension of $1024\times1024$ (Fig.~\ref{fig1}). The labelled images ($512\times512$ pixels) were used as training (80\%), testing (20\%), and evaluation (20\% of the training set) sets in the proposed U-Net networks.

\subsection{U-Net Model Architectures}
\begin{figure}[htbp]
\graphicspath{ {./images/} }
    \centering
   \includegraphics[width=1\textwidth]{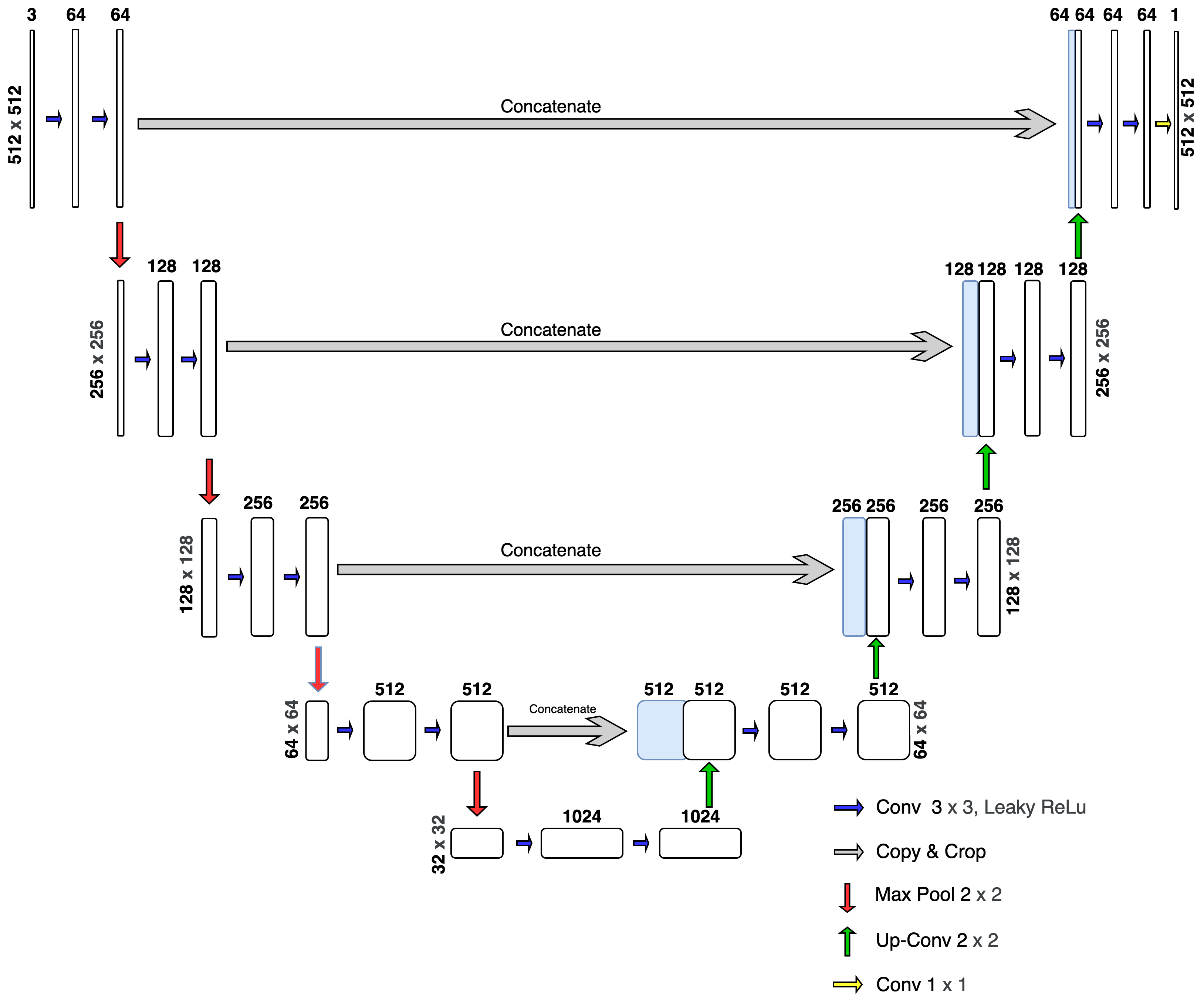}
    \caption{Architecture of the proposed simple U-Net model.} \label{fig2}
\end{figure}

The U-Net \cite{Ronneberger2015} is a semantic segmentation method proposed on the FCN architecture. The FCN consists of a typical encoder-decoder convolutional network. This architecture includes several feature channels to combine shallow and deep features. The deep features are used for positioning, whereas the shallow features are utilized for precise segmentation.The architecture of the simple U-Net was chosen (Fig.~\ref{fig2}) for training the model with the specific size of input images.

The first layer of the encoder part consists of the input layer, which accepts RGB images with the size $512\times512$. Each level in the five-"level" U-Net structure includes two 3$\times$3 convolutions. Batch normalization follows each convolution, and "LeakyReLu" activation functions follow a rectified linear unit. In the down-sampling (encoder) part (Fig.~\ref{fig2}, left part), each "level" in the encoder consists of a $2\times2$  max pooling operation with the stride of two.  The max-pooling process extracts the maximal value in the $2\times2$ area. By completing down-sampling in each level of the encoder part, convolutions will double the number of feature channels.

In the up-sampling (decoder) section (Fig.~\ref{fig2}, right part), the height and width of the existing feature maps are doubled in each level from bottom to top. 
Then, the high-resolution deep semantic and shallow features were combined and concatenated with the feature maps from the encoder section. After concatenation, the output feature maps have channels twice the size of the input feature maps. 
The output decoder layer at the top with a $1\times1$ convolution size predicts the probabilities of pixels. Padding in the convolution process allowed to achieve the same input and output layers size.
The computational result, combined with the Binary Focal Loss function, becomes the energy function of the U-Net.

\begin{figure}[htbp]
\graphicspath{ {./images/} }
    \centering
    \captionsetup{justification=centering}
   \includegraphics[width=\textwidth]{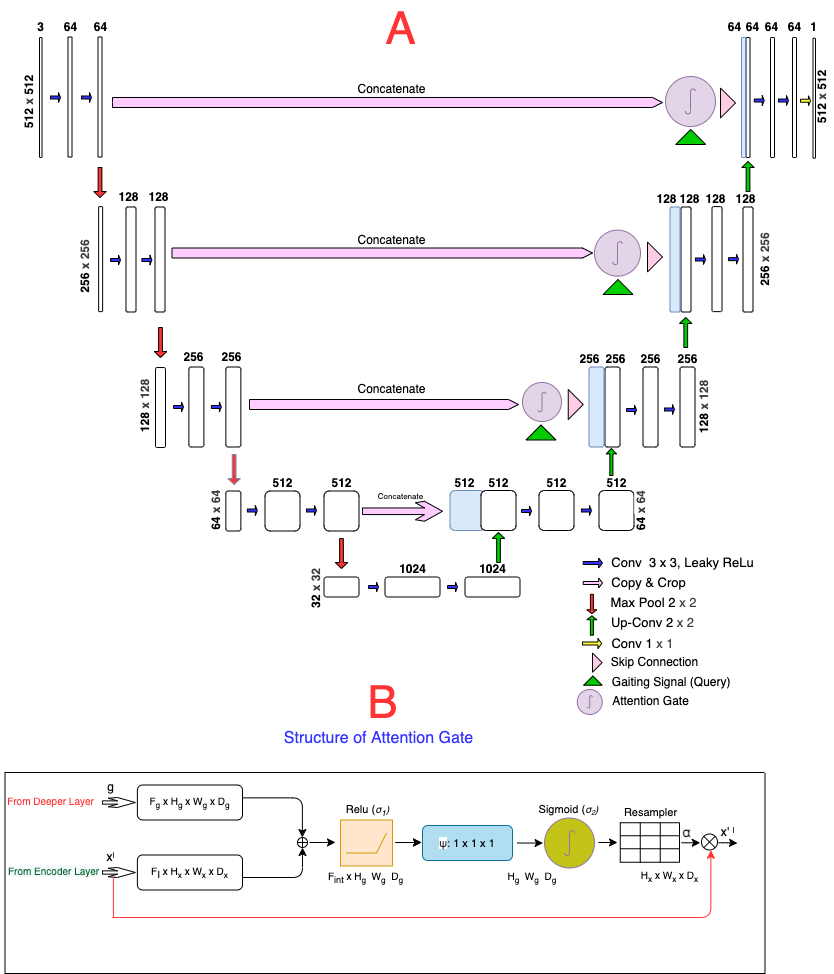}
    \caption{$A$) Architecture of the proposed Attention U-Net model, $B$) the attentive module mechanism. The size of each feature map is shown in $H\times W\times D$, where $H$, $W$, and $D$ indicate height, width, and number of channels, respectively.} \label{fig3}
\end{figure}
Between each Encoder-Decoder layer in the simple U-Net (Fig.~\ref{fig2}), there is a connection combining the down-sampling path with the up-sampling path to achieve the spatial information. Nevertheless, at the same time, this process brings also many irrelevant feature representations from the initial layers. The {self-attention} U-Net architecture (Fig.~\ref{fig3}-$A$) with an impressive performance in medical imaging \cite{Oktay2018} was applied to prevent this problem and improve semantic segmentation result achieved by standard U-Net. As an extension to the standard U-Net model architecture, the attention gate at the skip connections between encoder and decoder layers highlights the remarkable features and suppresses activations in the irrelevant regions. The advanced function of an attention mechanism is to map a set of key-value pairs and a query to an output. The key, query, values, and outputs are vectors. The compatibility function of the query, together with the corresponding key, is computed to be assigned by weights. Then, weighted sums of the values are computed and generate the output. The weights represent the relative importance of the inputs (the keys) for a particular output (the query) \cite{Vaswani2017}.
In this way, the attention gate improves the model sensitivity and performance without requiring complicated heuristics.

The attention gate (Fig.~\ref{fig3}-$B$) has two inputs: $x^l$ and $g$. Input $x^l$ comes from the skip connection from the encoder layers. Since coming from the early layers, input $x^l$ contains better spatial information. Providing $x^l$ is an output from layer $l$, a feature activation can be formulated as

\begin{equation}
x^l_i = \sigma_1(\sum_{c'\in F_1} x^{l-1}_{c'} \circledast k_{c',c}),
\end{equation}
by applying a rectified linear unit $\sigma_1(x^l_{i,c}) = \mbox{max}(0,x^l_{i,c})$ repeatedly, where $i$ and $c$ correspond to spacial and channel dimensions, respectively, and $F_1$ denotes the number of feature maps in layer $l$ and $\circledast$ indicates the convolution operation.

Input $g$ -- a gating signal -- comes from a deeper network layer and contains a better feature representation and contextual information to determining the focus region. Attention coefficients $\alpha \in [0,1]$ determine, extract, and preserve the valuable features corresponding to the important part of the image regions. The attention part weights different images' parts. This process will add the weights to the pixels based on their relevance in the training steps. The image's relevant parts will get higher weights than the less relevant parts. The output of the attention gate is the multiplication of the input feature maps $x^l_{i,c}$ and the achieved attention coefficient $\alpha$:

\begin{equation} \label{Eq0_0}
\mbox p^I_{att} = \psi^T(\sigma_1(W^T_x x^I_i+ W^T_g g_i + b_g)) + b_\psi,
\end{equation}

\begin{equation} \label{Eq0_1}
\alpha^I_i = \sigma_2(p^I_{att}(x^I_i, g_i; \Theta_{att})),
\end{equation}
where parameter $\sigma_2$ represents the sigmoid activation function and $\Theta_{att}$ contains parameters including linear transformations $W_x$ and $W_g$, function $\psi$ and bias terms $b_\psi$ and $b_g$ \cite{Oktay2018}. The achieved weights are also trained in the training process and make the trained model more attentive to the relevant regions.

\begin{figure}[htbp]
\graphicspath{ {./images/} }
    \centering
    \captionsetup{justification=centering}
    \includegraphics[width=\textwidth]{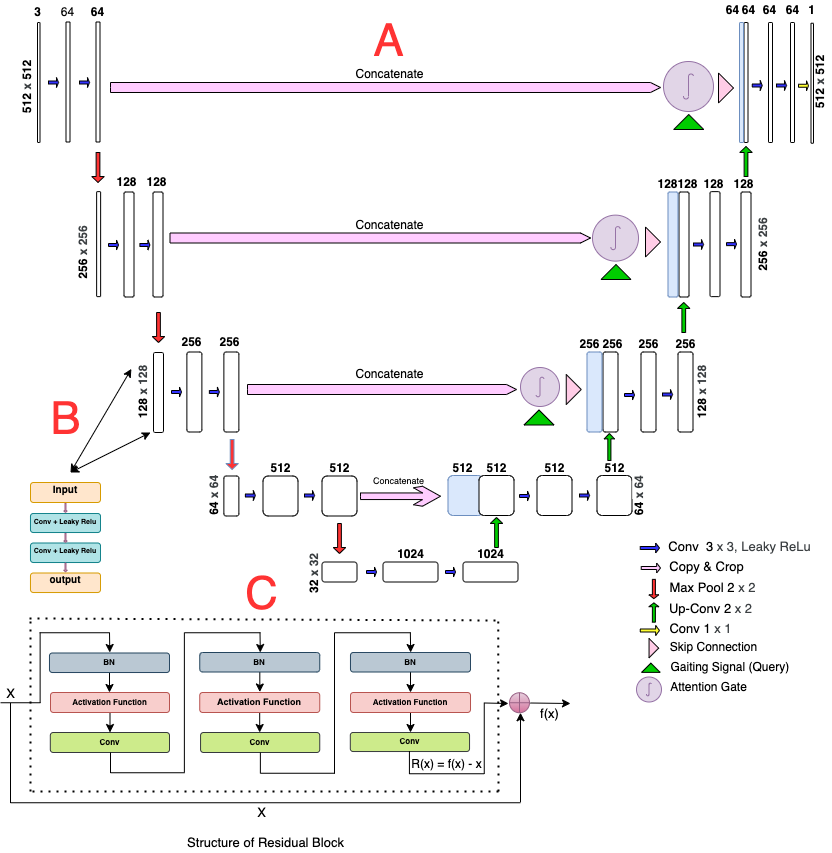}
    \caption{($A$) Architecture of the Residual Attention U-Net model. ($B$) Each U-Net layer structure. ($C$) The sample of residual block progress. $BN$ refers to Batch Normalization.} \label{fig4}
\end{figure}
Another architecture used in this study and developed based on the U-Net models (originally for nuclei segmentation \cite{Alom2018}) is the Residual U-Net. The simple U-Net architecture was built based on repetitive Convolutional blocks in each level (Fig.~\ref{fig4}-$B$). Each of these Convolutional blocks consists of the input, two steps of the convolution operation followed by the activation function and the output. On the other hand, we face the vanishing gradient problem when dealing with very deep convolutional networks. The residual step was applied to update the weights in each convolutional block incrementally and continuously (Fig.~\ref{fig4}-$C$) to enhance the U-Net architecture performance by overcoming the vanishing gradient problems.

In the traditional neural networks, each convolutional blocks feed the next blocks. The other problem in a DCNN-based network, such as stacking convolutional layers, is that a deeper structure of these kind of networks will affect generalization ability. To overtake this problem, the skip connections--the residual blocks--improve the network performance, with each layer feeding the next layer and layers about two or three steps apart (Fig.~\ref{fig4}--$C$). The Residual and Attention U-Net architecture were connected to build more effective and high-performance models from our datasets and improve segmentation results. 

\begin{table}[htbp]
\scriptsize
\centering
\caption{Number of the trainable parameters and the run time for each U-Net model.}
\label{tab:Exp_Time}
\begin{tabular}{ccc}
\hline
\textbf{Network}            & \textbf{Run time} & \textbf{Training parameter} \\ \hline
\textbf{U-Net}              & 3:42':18''           & 31,402,501                  \\
\textbf{Attention U-Net}    & 4:04':23''           & 34,334,665                  \\
\textbf{Residual Att U-Net} & 4:11':24''           & 39,090,377                  \\ \hline
    \end{tabular}
\end{table}

\begin{figure}[htbp]
\graphicspath{ {./images/} }
    \centering
    \captionsetup{justification=centering}
   \includegraphics[width=\textwidth]{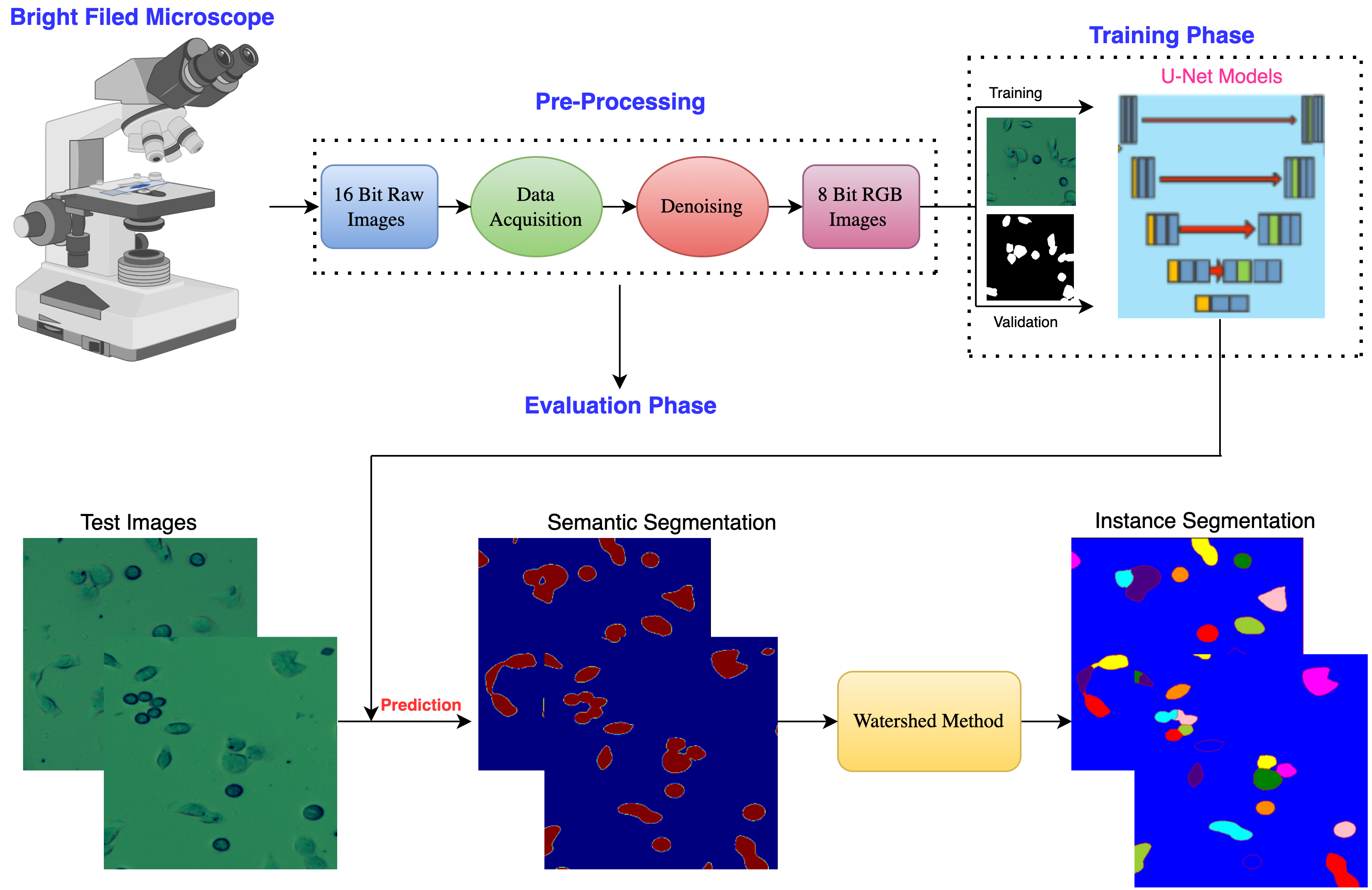}
    \caption{Flowchart of methodology applied in this study.} \label{fig5}
\end{figure}

The watershed algorithm based on morphological reconstruction \cite{Zhang2011} was applied after completion of the semantic segmentation by U-Net methods described above. The U-Net semantic segmentation results were first transformed into a binary image using the Otsu method \cite{Otsu1979}. After that, the background was determined using ten iterations of binary dilation. The simple Euclidean distance transform defined the foreground of eroded cell regions. The unknown region was achieved by subtraction of the particular foreground region from the background. The watershed method applied to the unknown regions separated the cell borders. The watershed segmentation further helped to solve the over- and under-segmented regions and specify each separated cell by, e.g., cell diameters, solidity, or mean intensity. The segmentation results were optimized using the marked images. Wrongly detected residual connections between different cell regions were cut off, which improved the method accuracy. Figure \ref{fig5} presents a general diagram of the proposed U-Net based methods. The U-Net models are hosted on the GitHub~\cite{GitHub}.

\subsection{Training Models}
The computation was implemented in Python 3.7. The framework for deep learning was Keras, and the backend was Tensorflow \cite{Abadi2015}. The whole method, including the Deep Learning framework, was transferred and executed on the Google Colab Pro account with P100 and T4 GPU, 24 Gb of RAM, and 2 vCPU \cite{GoogleColabPro}. After data preprocessing (Sect.~\ref{dataprep}), The primary dataset was divided into training (80\%) and test (20\%). A part (20\%) of the training set was used for model validation in the training process to avoid over-fitting and achieve higher performance. Among a 500-image dataset of the mixture of under-, over-, and focused images, 320 images were randomly selected to train the model, and 80 images were chosen randomly to validate the process. The rest of the 100 dataset images were considered for testing and evaluating the model after training.

Before the training, the images were normalized: the pixel values were rescaled in the range from 0 to 1. Since all designed network architectures work with a specific input image size, all datasets were resized to $512\times512$ pixel size. Data augmentation parameters were also applied in training all three U-Net architectures. The optimized values of the hyperparameters used in the training process are written in Tab. \ref{tab:Hyper_Param}. The "rotation range" represents an angle of the random rotation, "width shift range" represents an amplitude of the random horizontal offset, "height shift range" corresponds to an amplitude of the random vertical offset, "shear range" is a degree of the random shear transformation, "zoom range" represents a magnitude of the random scaling of the image. Early stopping hyperparameters were applied to avoid over-fitting during the model training.
The patient value was considered as 15. The activation function was set to the LeakyRelu, and the Batch size was set to 8. To optimize the network, we chose the Adam optimizer and set the learning rate to 10$^{-3}$. 

\begin{table}[htbp]
\scriptsize
  \centering
  \caption{Hyperparameters setting for all three U-Net models.}
  \label{tab:Hyper_Param}
   \begin{tabular}{@{}ll@{}}
   \toprule
\textbf{Parameter name} & \textbf{Value} \\ \midrule
Activation function     & LeakyRelu     \\ 
Learning rate           & 10$^{-3}$           \\ 
Batch size              & 8              \\ 
Epochs number           & 100            \\ 
Early stop              & 15             \\ 
Step per epoch          & 100            \\ 
Rotation range          & 90             \\ 
Width shift range        & 0.3            \\ 
Height shift range      & 0.3            \\ 
Shear range             & 0.5            \\ 
Zoom range              & 0.3            \\ \bottomrule
  \end{tabular}
\end{table}

Semantic image segmentation can be considered as a pixel classification as either the cell or background class. The Dice loss was used to compare the segmented cell image with the GT and minimize the difference between them as much as possible in the training process. One of the famous loss functions used for semantic segmentation is the Binary Focal Loss (Eq.~\ref{Eq1}) \cite{Lin2017}:

\begin{equation} \label{Eq1}
\mbox{Focal Loss} = -\alpha_t(1 - p_t)^\gamma \log(p_t),
\end{equation}
where $p_t \in [0, 1]$ is the model’s estimated probability for the GT class with label y = 1; a weighting factor $\alpha_t \in [0, 1]$ for class 1 and $1-\alpha_t$ for class $-1$; $\gamma \geq 0$ is a tunable focusing parameter. The focal loss can be enhanced by the contribution of hardly segmented regions (e.g., cells with vanished borders) and distinguish parts between the background and the cells with unclear borders. The second benefit of the focal loss is that it controls and limits the contribution of the easily segmented pixel regions (e.g., sharp and apparent cells) in the image at the loss of the model. In the final step, updating the gradient direction is under the control of the model algorithm, dependent on the loss of the model.

\subsection{Evaluation metrics} \label{Evaluation metrics}
The proposed semantic segmentation models were evaluated by different metrics (Eqs. \ref{Eq2}--\ref{Eq6}), where TP, FP, FN, and TN are true positive, false positive, false negative, and true negative metrics, respectively \cite{Pan2017}. The metrics were computed for all test sets and explained as mean values (Tab. \ref{tab:Exp_Res}).

Overall pixel accuracy (Acc) represents a per cent of image pixels belonging to the correctly segmented cells. Precision (Pre) is a proportion of the cell pixels in the segmentation results that match the GT. 
The Recall (Recl) represents the proportion of cell pixels in the GT correctly identified through the segmentation process. 
The F1-score or Dice similarity coefficient states how the predicted segmented region matches the GT in location and level of details and considers each class's false alarm and missed value. This metric determines the accuracy of the segmentation boundaries \cite{Csurka2013} and have a higher priority than the Acc. Another 
essential evaluation metric for semantic image segmentation is the Jaccard similarity index known as Intersection over Union (IoU). This metric is a correlation among the prediction and GT \cite{Long2015,Vijay2015}, and represents the overlap and union area ratio for the predicted and GT segmentation.

\begin{equation} \label{Eq2}
\mbox{Acc} =\frac{\mbox{Correctly Predicted Pixels}}{\mbox{Total Number of Image Pixels}} = \frac{\mbox{TP + TN}}{\mbox{TP + FP + FN + TN}}
\end{equation}

\begin{equation}
\mbox{Pre} = \frac{\mbox{Correctly Predicted Cell Pixels}}{\mbox{Total Number of Predicted Cell Pixels}} = \frac{\mbox{TP}}{\mbox{TP + FP}}
\end{equation}

\begin{equation}
\mbox{Recl} =\frac{\mbox{Correctly Predicted Cell Pixels}}{\mbox{Total Number of Actual Cell Pixels}} = \frac{\mbox{TP}}{\mbox{TP + FN}}
\end{equation}

\begin{equation}
\mbox{Dice} =\frac{\mbox{2 $\times$ Pre $\times$ Recl}}{\mbox{Pre + Recl}} = \frac{\mbox{2 $\times$ TP}}{\mbox{2 $\times$ TP + FP + FN}}
\end{equation}

\begin{equation} \label{Eq6}
\mbox{IoU} = \frac{\mid y_t \cap y_p \mid}{\mid y_t \mid + \mid y_p \mid - \mid y_t \cap y_p \mid} = \frac{\mbox{TP}}{\mbox{TP + FP + FN}}
\end{equation}



\begin{sidewaysfigure}
\graphicspath{ {./images/} }
   \includegraphics[width=1\textwidth]{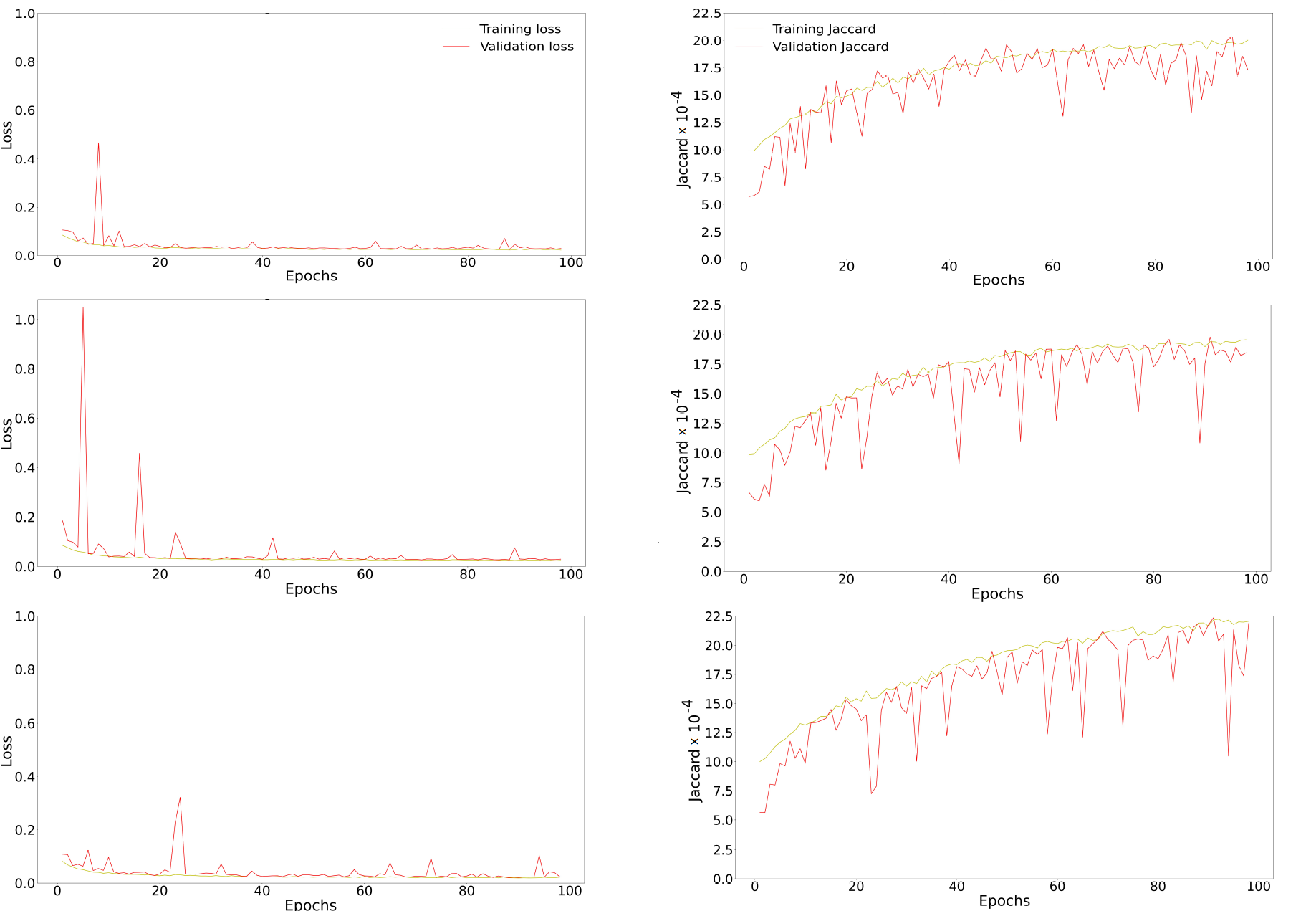}
   \captionsetup{justification=centering}
    \caption{Training/validation plots for Simple U-Net (left column), Attention U-Net (middle column), and Residual Attention U-Net (right column).} \label{fig6}
\end{sidewaysfigure}

\section{Results}
All three models were well trained and converged after running 100 epochs based on training/validation loss and Jaccard plots per epochs (Fig.~\ref{fig6}). The hyperparameter values listed in Table~\ref{tab:Hyper_Param} were selected to tune for the best training performance and stability. Then, the test datasets were used to evaluating the achieved models. All trained models were assessed (Tab.~\ref{tab:Exp_Res}) using the metrics in Eqs.~\ref{Eq2}--\ref{Eq6}.

\begin{sidewaysfigure}
\graphicspath{ {./images/} }
    \centering
   \captionsetup{justification=centering}
   \includegraphics[width=1.0\textwidth]{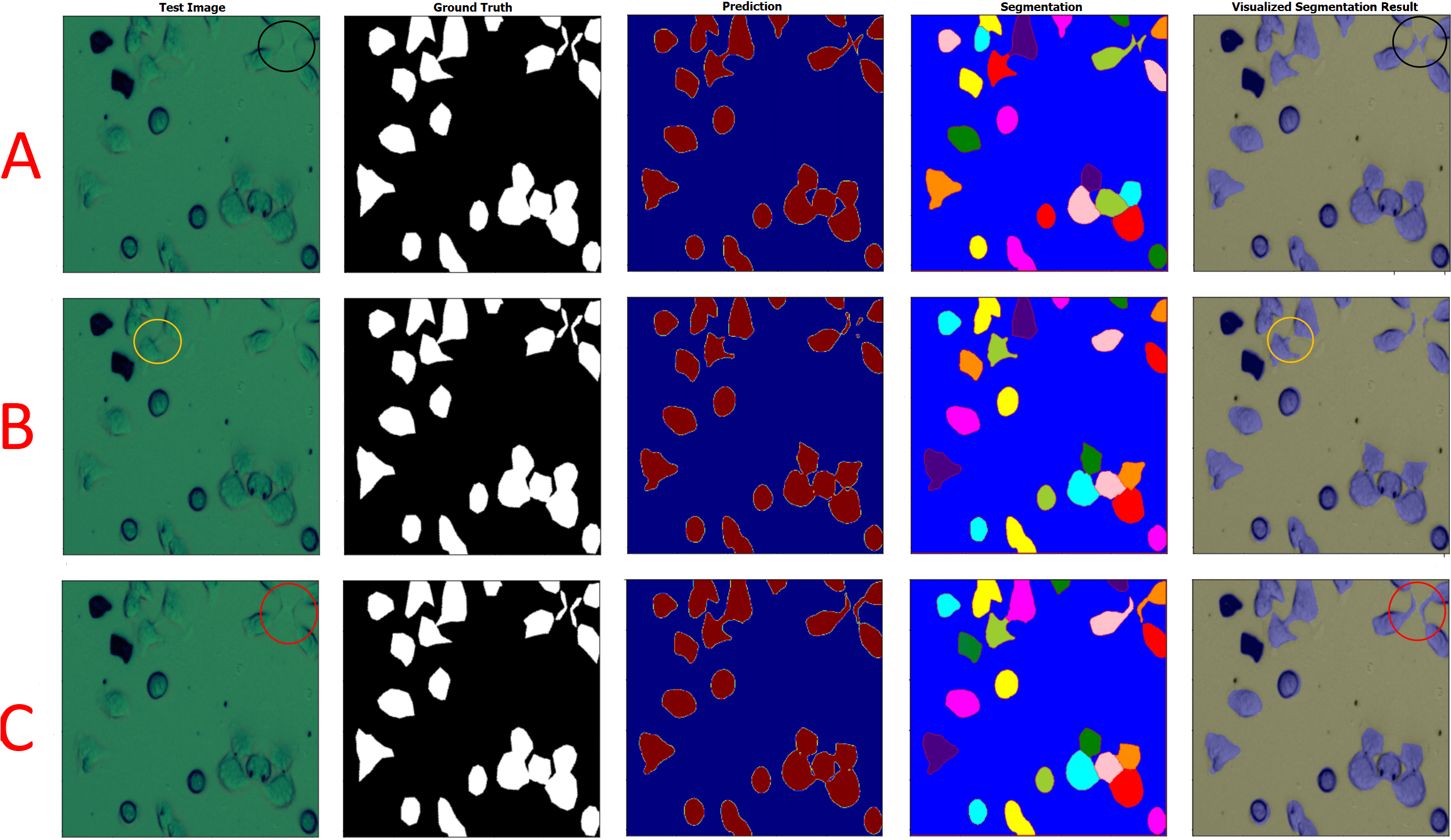}
    \caption{Segmentation results for $A$) the simple U-Net (the black circle highlights the non-segmented, vanished cell borders), $B$) Attention U-Net (the yellow circle highlights the undersegmentation problem), and $C$) the Residual Attention U-Net (red circle shows the successful segmentation of the cell borders). The image size is $512 \times 512$.} \label{fig7}
\end{sidewaysfigure}

Training the model with the simple U-Net method took the shortest run time with the lowest trainable number of parameters (Tab.~\ref{tab:Exp_Time}). Compared with the Attention U-Net and Residual Attention U-Net, the run time difference is not huge in terms of increasing trainable parameters. The computational cost also did not increase dramatically compared with the acceptable improvement in the model performance. Figure \ref{fig7} presents the segmentation results achieved by three different U-Net models. The simple U-Net segmentation result did not distinguish some vanished cell borders (Fig. \ref{fig7}--$A$, black circle). The Attention U-Net (Fig. \ref{fig7}--$B$) detected cells with the vanish borders more efficiently than the simple U-Net. However, the Attention U-Net segmentation suffers from under-segmentation in some regions (visualized by the yellow circle). The outcome of the Residual Attention U-Net method (Fig. \ref{fig7}--$C$, red circle) achieved more accurate segmentation of the vanished cell borders. The watershed binary segmentation after the Residual Attention U-Net networks separated and identified the cells with the highest performance (Fig. \ref{fig7}).

As seen in Mean-IoU, Mean-Dice, and Accuracy metrics (Tab.~\ref{tab:Exp_Res}), the Attention U-Net model showed better segmentation performance than the simple U-Net model in the same situation. The segmentation results were further slightly improved after applying the residual step into the Attention U-Net.

\begin{table}[htbp]
\scriptsize
   \centering
   \captionsetup{justification=centering}
   \caption{Results for metrics evaluating the U-Net Models. Green values represent the highest segmentation accuracy for the related metric.}
   \captionsetup{justification=centering}
   \label{tab:Exp_Res}
   \begin{adjustbox}{width=\textwidth}
   \begin{tabular}{ccccccc}
\hline
\multicolumn{1}{c}{\textbf{Network}} & \textbf{Accuracy} & \textbf{Precision} & \textbf{Recall} & \textbf{m-IoU}    & \textbf{m-Dice}    \\ \hline
\textbf{U-Net}                        & 0.957418          & 0.988269           & 0.961264        & 0.950501         &       0.974481     \\ 
\textbf{Attention U-Net}              & 0.959448          & 0.985663           & 0.965736        & 0.952471         &          0.975511  \\ 
\textbf{Residual Att U-Net}           & 0.960010          & 0.986510           & 0.965574        & \cellcolor{green!10}0.953085 & \cellcolor{green!10}0.975840 \\ \hline
     \end{tabular}
  \end{adjustbox}
\end{table}

\section{Discussion}
The analysis of bright-field microscopy image sequences is challenging due to living cells' complexity and temporal behaviour. We have to face (1) irregular shapes of the cells, (2) very different sizes of the cells, (3) noise blobs and artefacts, and (4) vast sizes of the time-lapse datasets. Traditional machine learning methods, including random forests and support vector machines, cannot deal with some of these difficulties in terms of higher computational cost and longer run time for huge time-lapse datasets. The traditional methods suffer from low performance in vanishing and tight cell detection and segmentation and are sensitive to training steps \cite{SommerC2011,Tikkanen2015}. The DL methods have been rapidly developed to overcome these problems. The U-Net is one of the most effective semantic segmentation methods for microscopy and biomedical images \cite{Ronneberger2015}. This method is based on the FCN architecture and consists of encoder and decoder parts with many convolution layers.

The image data used to train the Residual Attention model are specific in the way of acquisition. Firstly, the optical path was calibrated to obtain the number of photons that reaches each camera pixel with increasing illumination light intensity. This gave a calibration curve (image pixel intensity vs the number of photons reaching the relevant camera pixel) to correct the digital image pixel intensity. This step ensured homogeneity in digital image intensities to improve the quality of cell segmentation by the neural networks. We work with the low-compressed telecentric transmitted light bright-field high-pixel microscopy images. The bright-field light microscope allows us to observe living cells in their most natural state. Due to the object-sided telecentric objective, the final digital raw image of the observed cells is high-resolved and low-distorted, with no light interference halos around objects.

The procedure compressed the raw colour images to ensure the least information loss at the quarter-pixel-resolution decrease of the image. The final pixel resolution of the images inputting into the neural network is higher (512$\times$512) than in the case of any other neural network datasets. By preserving high image resolution as much as possible, the demands on the neural network's computational memory and performance parameters were increased. 

As our microscope and acquired microscopy data are unique, and were not used before in similar research, it is hard to compare the results with other works. Despite this, the performances of the proposed U-Net-based models were compared with similar microscopy and medical works (Tab.~\ref{tab:comparision}). Our first model was based on a simple U-Net structure and achieved the Mean-IoU score of 0.9505. We assume that better value of the Mean-IoU will be achieved after the hyperparameter optimization (Tab.~\ref{tab:Hyper_Param}). Ronnenberger et al. \cite{Ronneberger2015} achieved 0.920 and 0.775 Mean-IoU scores for U373 cell line in phase-contrast microscopy and HeLa cell line in Nomarski contrast, respectively. Pan et al. \cite{Pan2019} segmented nuclei from medical, pathological MOD datasets with 0.7608 segmentation IoU accuracy score using the U-Net.

We further implemented an attention gate into the U-Net structure (so-called Attention U-Net) to further improve the U-Net model performance by weighing the relevant part of the image pixels containing the target object. In this way, the Mean-IoU metric was improved to 0.9524. The achieved IoU score represents a noticeable improvement in the trained model performance compared with the simple U-Net model. To the best of our knowledge, not many researchers have applied the Attention U-Net to microscopy datasets, but recent papers are prevalently about its application to medical datasets. Microscopy and medical datasets have their complexity and structure, complicating the comparison of the method performances. Applying the Attention U-Net, pancreas \cite{Oktay2018} and liver tumour \cite{Wang2021} medical datasets showed 0.840 and 0.948 Dice metric segmentation accuracy, respectively.

\begin{table}[htbp]
\scriptsize
    \centering
    \captionsetup{justification=centering}
      \caption{Performances of the proposed networks and other networks proposed for microscopy and medical applications. Green highlighted value represent the highest segmentation accuracy in term of mentioned metric.}
    \label{tab:comparision}
     \begin{tabular}{cccc}
\hline
\textbf{Models}                                  & \textbf{IoU} & \textbf{Dice} & \textbf{Acc} \\ \hline
\textbf{proposed U-Net}                               & 0.9505       & 0.9744        & 0.9574       \\ 
\textbf{proposed Att U-Net}                           & 0.9524       & 0.9755        & 0.9594       \\ 
\textbf{proposed ResAtt U-Net}                     & \cellcolor{green!10}0.9530       & \cellcolor{green!10}0.9758        &\cellcolor{green!10} 0.9600       \\ 
U-Net \cite{Ronneberger2015}         & 0.9203       & 0.9019        & 0.9554       \\ 
U-Net \cite{Pan2019}                     & 0.7608       & -             & 0.9235       \\ 
U-Net+ \cite{Long2020}         & 0.567       & -        & -       \\ 
DenseNet \cite{Bagyaraj2021}         & -       & 0.911        & -       \\ 
SegNet \cite{Pan2019}                    & 0.7540       & -             & 0.9225       \\ 
Attention U-Net \cite{Oktay2018}             & -            & 0.840         & 0.9734       \\ 
Residual Attention U-Net \cite{Wang2021} & -            & 0.9081        & 0.9557       \\ 
Residual  U-Net \cite{patel2019} & -            & 0.8366        & -       \\ 
Residual Attention U-Net \cite{Qiangguo2020}      &    -          & 0.9655        & 0.9887       \\ \hline
\end{tabular}
\end{table}

The proposed model performance were improved by one step and obtained the Residual Attention U-Net to overcome the vanishing gradient problem and generalization ability. As a result, the segmentation accuracy  was slightly improved by reaching the Mean-IoU of 0.953. The Residual Attention U-Net showed the Dice coefficient of 0.9655 in the testing phase of medical image segmentation \cite{Qiangguo2020}. The Recurrent Residual U-Net (R2U-Net) achieved the Dice coefficient of 0.9215 in the testing phase of nuclei segmentation \cite{Alom2018}. Patel et al. \cite{patel2019} applied the Residual U-Net to bright-field absorbance image and achieved the Mean-Dice coefficient score of 0.8366.  Long et al. \cite{Long2020} applied the enhanced U-Net (U-Net+) to bright-field, dark-field, and fluorescence microscopy images and achieved the Mean-IoU score of 0.567. The U-Net with a dense convolutional network (DenseNet) was applied to detect and segment brain tumour cells  \cite{Bagyaraj2021} with the Dice score of 0.911 and the Jaccard index of 0.839.
 
\section{Conclusion}
Microscopy image analysis via deep learning methods can be a convenient solution due to the complexity and variability of this kind of data. This research aimed to detect and segment living human HeLa cells in images acquired using an original custom-made bright-field transmitted light microscope. Three types of deep learning U-Net architectures were involved in this research: the simple U-Net, Attention U-Net, and Residual Attention U-Net. 
The simple U-Net (Tab.~\ref{tab:Exp_Time}) has the fastest training time. On the other hand, the Residual Attention U-Net architecture achieved the best segmentation performance (Tab.~\ref{tab:Exp_Res}) with a run time slightly higher than the other two U-Net models.  

The Attention U-Net is a method to highlight only the relevant activations during the training process. This method can reduce the computational resource waste on irrelevant activations to generate more efficient models. 
The best segmentation performance was achieved due to the integration of the residual learning structure (to overcome the gradient vanishing) together with the attention gate mechanism (to integrate a low and high-level feature representation) into the U-Net architecture.
After extracting semantic segmentation binary results (Tab.~ \ref{tab:Exp_Res}), the watershed segmentation method was applied to separate the cells from each other, avoid over-segmentation, label the cells individually, and extract vital information about the cells (e.g., the total number of the segmented cells, cell equivalent diameter, mean intensity and solidity). Nevertheless, future works are still essential to expand the knowledge on multi-class semantic segmentation with different and efficient CNN's architecture and combine the constructed CNN models in the prediction process to achieve the most accurate segmentation result.

\section*{FUNDING}
This work was supported by the Ministry of Education, Youth and Sports of the Czech Republic -- project CENAKVA (LM2018099), by the European Regional Development Fund in frame of the project ImageHeadstart (ATCZ215) in the Interreg V-A Austria–Czech Republic programme, and by the project GAJU 017/2016/Z.

\section*{DECLARATION OF COMPETITING INTEREST}
The authors declare no conflict of interest, or known competing financial interests, or personal relationships that could have appeared to influence the work reported in this paper.

\section*{ACKNOWLEDGEMENT}
The authors would like to thanks our lab colleagues Šárka Beranová and Pavlína Tláskalová (both from the ICS USB) and Mohammad Mehdi Ziaei for their support of this study.

\section*{DATA AND CODE AVAILABILITY}

The U-Net models are hosted on the GitHub~\cite{GitHub} and other data on~the Dryad~\cite{Dryad}.



\bibliographystyle{elsarticle-num}
\bibliography{Ghaznavi}
\end{document}